# Interlaboratory study of ice adhesion using different techniques

Sigrid Rønneberg[1], Yizhi Zhuo[1], Caroline Laforte[2], Jianying He[1], Zhiliang Zhang[1]*

[1] Department of Structural Engineering, Norwegian University for Science and Technology (NTNU), NO-7491 Trondheim, Norway

[2] Anti-Icing Materials International Laboratory (AMIL), Université du Québec à Chicoutimi, 555 Blvd. de l'Université, Chicoutimi, Québec G7H 2B1, Canada

* Corresponding author. E-mail: zhiliang.zhang@ntnu.no. Telephone: +4773592530 / +4793001979

## Abstract

Low ice adhesion surfaces are a promising anti-icing strategy. However, reported ice adhesion strengths cannot be directly compared between research groups. This study compares results obtained from testing the ice adhesion strength on the same surface at two different laboratories, testing two different types of ice with different ice adhesion test methods at temperatures of -10°C and -18°C. One laboratory uses the centrifuge adhesion test and tests precipitation ice and bulk water ice, while the other laboratory uses a vertical shear test and tests only bulk water ice. The surfaces tested were bare aluminum and a commercial icephobic coating, with all samples prepared in the same manner. The results showed comparability in the general trends, surprisingly, with the greatest differences for bare aluminum surfaces at temperature -10°C. For bulk water ice, the vertical shear test resulted in systematically higher ice adhesion strength than the centrifugal adhesion test. The standard deviation depends on the surface type and seems to scale with the absolute value of the ice adhesion strength. The experiments capture the overall trends in which the ice adhesion strength surprisingly decreases from -10°C to -18°C for aluminum and is almost independent of temperature for a commercial icephobic coating. In addition, the study captures similar trends in the effect of ice type on ice adhesion strength as previously reported and substantiates that ice formation is a key parameter for ice adhesion mechanisms.

## Nomenclature

AMIL    – Anti-icing Materials International Laboratory
$A$    – ice-solid contact area
ARF    – Adhesion reduction factor
BWI    – Bulk water ice
CAT    – Centrifuge adhesion test
$F$    – Centrifugal force
IC    – Icephobic coating
MVD    – Median volume drop diameter
$m_{ice}$    – Mass of detached ice
NTNU    – Norwegian University of Science and Technology
PI    – Precipitation ice
$r$    – Radius of the beam at the center of mass for the ice sample
$\tau$    – Ice adhesion strength
VST    – Vertical shear test
$\omega$    – Angular velocity at ice detachment

# Introduction

Anti-icing surfaces, or icephobic surfaces, are a promising technique for passive ice removal and may help mitigate and avoid dangerous situations and unwanted icing in our daily life [1-4]. The most promising strategy for anti-icing surfaces is low ice adhesion surfaces, where the ice automatically detaches from the surface by its own weight or natural forces [5-7]. However, although the amount of research on low ice adhesion surfaces has steadily increased over the past few years [8] and record low ice adhesion strengths of below 1 kPa has been reported [9-11], each research group develops its own custom-built set-up for measuring ice adhesion strength [9, 12-15]. As a result, reported ice adhesion strength measurements cannot be directly compared [7, 8, 16, 17].

In this experimental study, the research groups at the Anti-icing International Materials Laboratory (AMIL) at the University of Québec in Chicoutimi and the Nanomechanical Lab at the Norwegian University of Science and Technology (NTNU) collaborate to compare obtained ice adhesion strength measurements. Both have laboratory facilities able to measure internally comparable ice adhesion strength in controlled environments. At AMIL, the ice adhesion strength is measured with a centrifuge adhesion test (CAT), which is illustrated in Figure 2. This centrifuge test is one of the most repeatable ice adhesion tests, although it cannot produce stress-strain curves [8, 17, 18]. For larger facilities, the CAT is a common way to measure ice adhesion strength, often for impact ice types produced with a freezing drizzle or in-flight icing simulation [19-31]. At NTNU, the ice adhesion strength is measured with a vertical shear test (VST), as illustrated in Figure 3. The VST is very common due to its simple and economical set-up and performance, although the location of the force probe impacts the ice adhesion strength greatly [32], and the stress distribution may not be completely uniform [8, 17, 18]. The VST is commonly in use by several research groups [7, 11, 32-39], and has been attempted as a standard for ice adhesion measurement utilizing only commercially available instruments [14].

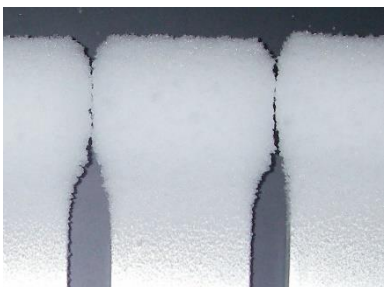
*Figure 1a Illustration of precipitation ice (PI) created at AMIL ($T_{air}$=-18°C).*

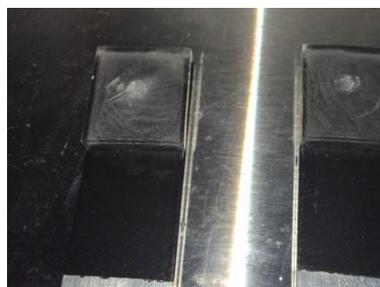
*Figure 1b Illustration of bulk water ice (BWI) created at AMIL ($T_{air}$=-18°C).*

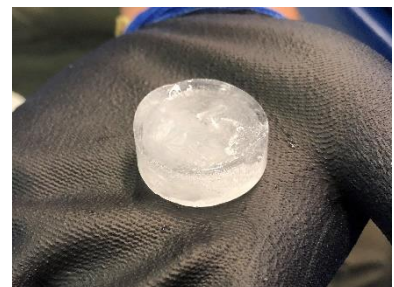
*Figure 1c Illustration of bulk water ice (BWI) created at NTNU ($T_{air}$=-18°C).*

When comparing reported ice adhesion strengths, it is also necessary to include the type of ice tested. Measured ice adhesion strength is highly dependent on the ice tested [40], and it is essential to test ice adhesion strength with a realistic ice type for low ice adhesion surfaces with a specific application in mind. In this study, both ice from freezing precipitation and ice from bulk water samples are tested, see Figure 1. These ice types are analogous to those presented elsewhere [40], and while precipitation ice (PI) is a form of ice from impacting freezing supercooled droplets (Figure 1a), bulk water ice (BWI) is a static, non-impact type of ice (Figure 1b and Figure 1c). BWI is the most common ice for testing of ice adhesion strength [5, 9, 10, 12, 33, 34, 41-50], although PI is also previously studied [19, 24, 51, 52]. For most practical applications, PI is more realistic than BWI [8, 17].

The comparison of ice adhesion strength measured at the facilities of AMIL and NTNU for the two types of ice showed that all results are comparable within the general trends between NTNU and AMIL, with the greatest differences for aluminum surfaces at -10°C. However, there are considerable differences between different laboratories. The study provides further evidence that the ice formation is a key parameter in predicting the ice adhesion on different surfaces.

# Experimental details

The ice adhesion strength of two surfaces were tested by both AMIL and NTNU in their respective facilities. The surfaces tested were bare aluminum 6061-T6, and aluminum covered with EC-3100, a two component, water-based, icephobic, non-stick coating from Ecological Coating, LLC. The testing of these surfaces has been reported previously [51, 53, 54]. The aluminum samples were polished with Walter BLENDEX Drum fine 0724 M4. To ensure similar surfaces, all the tested surfaces were produced at AMIL facilities and transported to NTNU for testing. Each surface was tested only once to discount the durability aspect of the surfaces. All ice was generated with demineralized water of resistivity 18 MΩcm. Both temperatures of -10°C and -18°C were tested, with six different samples from each configuration to generate average ice adhesion strength. Full experimental protocol is available as part of supplementary materials.

## AMIL facility

The samples tested at AMIL were in the form of bars fit to the CAT apparatus, with the iced area on one side and a counterweight on the other. The bars had length 340 mm and thickness 6.3 mm, with icing occurring over an area of about 1100 mm². This area was measured more precisely after the ice adhesion test in order to have the exact ice-surface detached surfaces.

PI was created through a freezing drizzle in a cold room of constant temperature and a relative humidity of 80% ± 2%. Six samples were iced simultaneously, with water of a median volume drop diameter (MVD) of 324 μm and an initial temperature of 4°C at the exit of the sprayer nozzle. The surfaces had initial temperature of the testing temperature, meaning either -10°C or -18°C. As the water hits the sample surface, it has become supercooled and freezes on contact. Water impact speed is due to gravity as the water droplets fall from the nozzle, and is estimated to about 5 ms$^{-1}$. The samples were iced for 33 minutes and kept in the cold room for 1 hour between icing and ice adhesion test to allow the ice to thermally stabilize.

BWI was created in the same cold room by freezing water in silicon molds from MoldMax30 by Smooth-On [55]. The silicon molds had the same dimensions as the area iced during the freezing drizzle, to generate ice samples as similar as possible to the PI. The molds were filled full of water, with the samples placed on top of the molds in contact with the water for freezing to occur. The surfaces and water were at room temperature at the start of the icing. Freezing time was 3 hours, after which the molds were removed. The ice adhesion test was conducted after 15 minutes, in which the samples were weighed and measured.

The ice adhesion strength was measured with the CAT apparatus [51], see Figure 2. The CAT apparatus consists of a centrifuge, a placed sample beam, a counterweight to stabilize the bar with the ice sample, and a cover. The apparatus is placed within the cold room, ensuring in situ measurements of the ice adhesion strength for PI and BWI. The balanced and iced sample bars were spun in the centrifuge at an accelerating speed of 300 rpms$^{-1}$ until the ice was detached by the centrifugal force. Piezoelectric cells situated around the cover instantly detected the detachment of the ice, giving a detachment angular velocity. The ice adhesion strength is calculated as the centrifugal shear stress at the position of the center of mass of the ice sample at detachment divided by the ice-solid contact area [51].

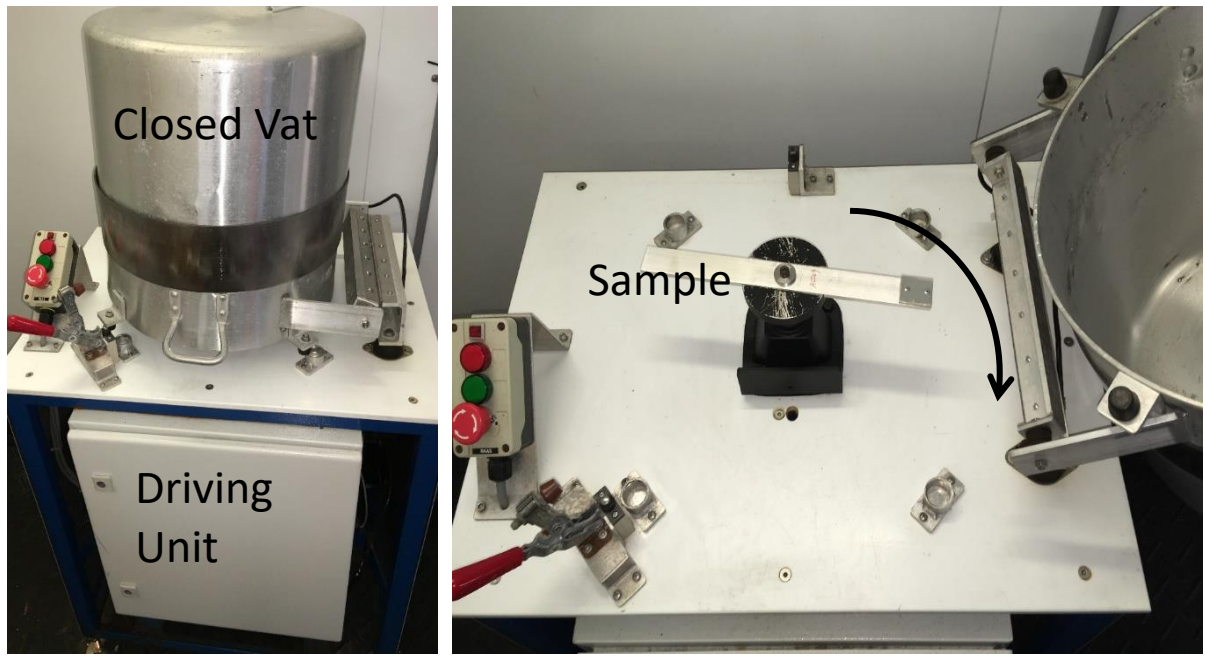

*Figure 2 AMIL CAT apparatus.*

## NTNU facility

The surfaces tested at NTNU were approximately square surfaces of width 7.3 cm, height 7.2 cm and thickness 25 mm. The ice sample was frozen in the middle of the surface for testing. Both water and surfaces where initially at room temperature for the testing at NTNU.

The ice tested at NTNU was BWI. For temperature of -18°C, the ice samples were frozen in a freezer, while for temperature of -10°C, the ice was frozen in a cold room situated at a slight distance from the ice adhesion test. For both temperatures, the ice was frozen ex situ, and required transportation through room temperature to the testing rig where the samples were again placed in the original temperature for ice adhesion tests. For temperature -18°C, the transport time was about one minutes and 30 seconds, while for temperature -10°C, the transport time was about three minutes. To account for the transport from the cold room, the samples were transported in a box made of expanded polystyrene with freezer elements. Both the box and freezer elements were placed in the cold room for thermal equilibration before and after the transportation. After the transportation, the ice samples were placed in the ice adhesion test chamber for 15 minutes before testing to achieve thermal stability.

The BWI samples were frozen on the tested surfaces in a polypropylene centrifuge tube mold with wall thickness of 1 mm and inner diameter of 27.5 mm. Silicone grease [56] was used to fasten the tube mold to the tested surface to avoid leakage during water insertion. 5 mL of deionized water was inserted into the mold with a syringe to avoid air at the ice-solid interface, and pressure from a 200 g metal cylinder was placed on top of the tube to avoid water leakage during freezing. The water was frozen for 3 hours before it was moved to the testing apparatus.

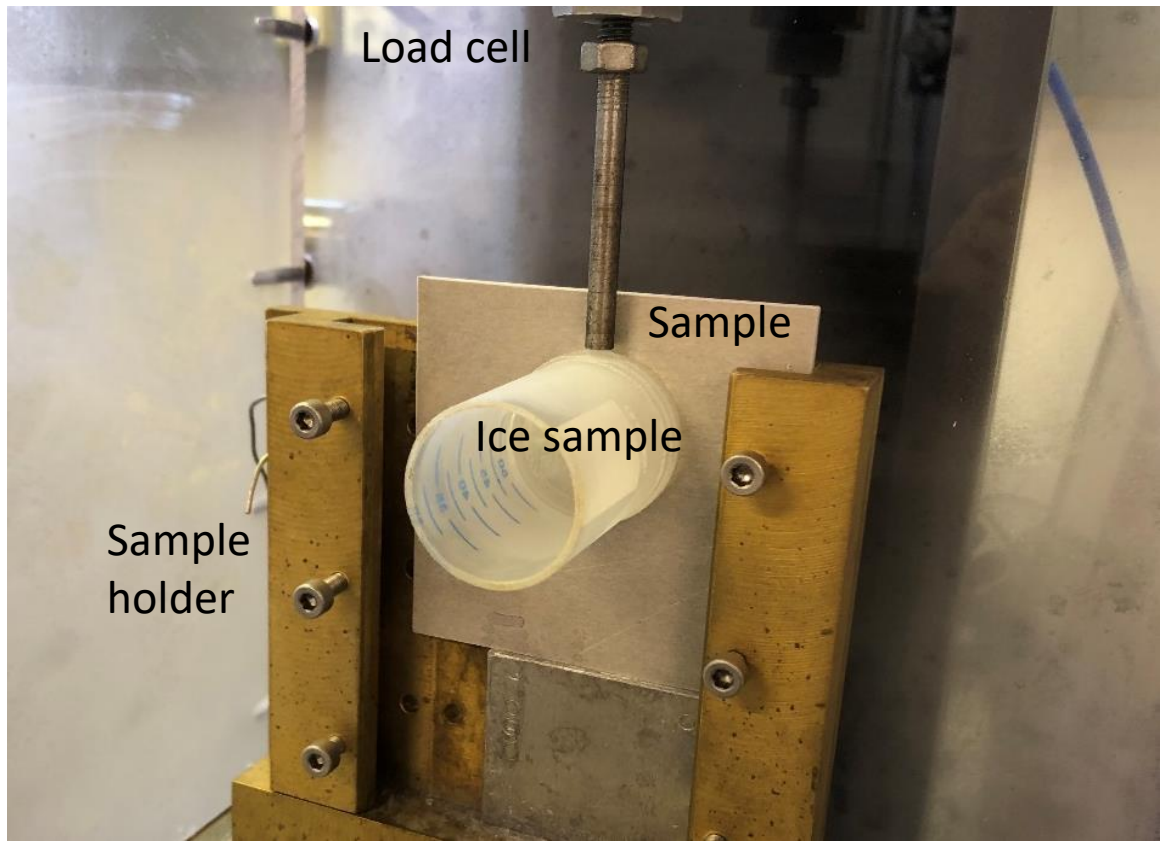

*Figure 3 VST as utilized at NTNU to measure ice adhesion strength.*

The ice adhesion test was performed with a VST and a custom-built set-up as modelled from other facilities [14], see Figure 3. The detachment force was measured with an Instron machine (model 5944) with load cell capacity of 2 kN (2530 Series static load cells), equipped with a home-built cooling system and chamber. The force probe fixed to the load cell was 5 mm in diameter and imposed an increasing force on the tube-encased ice samples with an impact velocity of 0.01 mms$^{-1}$. The placement of the probe was at the same point on the sample each test, situated 3 mm away from the tested surface during loading. The loading curve was recorded, and the peak value of the shear force was divided by the contact area to obtain the ice adhesion strength. As the probe distance is small and the measured ice adhesion strength is above 10 kPa for all tests, gravity can be discarded as negligible [8].

## Results and discussion

The measured ice adhesion strengths are shown in Figure 4. It can be seen that all results are comparable to a degree, with the greatest differences for aluminum surfaces at -10°C. Table 1 presents an overview of the ice adhesion and standard deviation obtained from both laboratories. To obtain an average value, six different samples were tested at AMIL, except for BWI on aluminum at -10°C where only four samples could be tested. At NTNU, averages were created from five samples. All the data are given in the supplementary materials.

From Figure 4, it may be seen that for BWI, the NTNU VST method systematically yields higher ice adhesion strength than AMIL CAT method for both aluminum surfaces and the icephobic coating. However, the standard deviation depends on the surface type. For bare aluminum, the deviation for VST is higher than CAT, while for the icephobic coating, the opposite trend is observed.

When comparing the two surface types for all ice types, all tests show larger error bars for aluminum than for the icephobic coating. This high standard deviation is in accordance with other studies of ice adhesion strength, and may be an inherent property of the ice removal mechanisms [8, 57]. The ice adhesion strengths for the icephobic coating from both laboratories are close to each other, but shows larger variations for BWI, up to 58% compared to up to only 18% for PI.

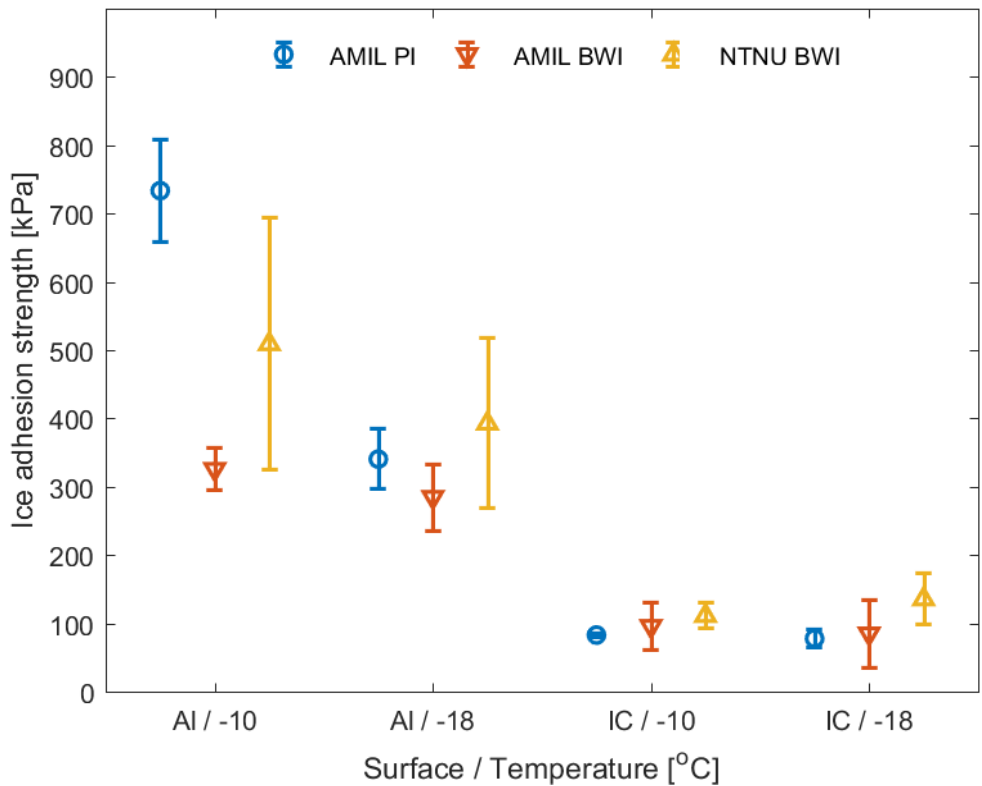

*Figure 4 Measured ice adhesion strengths. Aluminum surfaces as denoted as Al, while the surfaces with icephobic coating are denoted IC. All three ice types created are shown for each surface-temperature combination.*

*Table 1 Overview of mean values and standard deviations of ice adhesion strength.*

| Surface / Temperature | Ice adhesion strength [kPa ± SD (%)] | | |
|---|---|---|---|
| | **AMIL PI** | **AMIL BWI** | **NTNU BWI** |
| **Aluminum / -10°C** | 734 ± 75 (10%) | 326 ± 30 (9%) | 509 ± 185 (36%) |
| **Aluminum / -18°C** | 340± 44 (13%) | 285 ± 49 (17%) | 393 ± 124 (32%) |
| **Coating / -10°C** | 83 ± 3 (4%) | 96 ± 34 (35%) | 111 ± 19 (17%) |
| **Coating / -18°C** | 78 ± 14 (18%) | 85 ± 49 (58%) | 135 ± 38 (28%) |

The effect of decreasing temperature varied for the tested surfaces. At AMIL, there was a marked decrease of ice adhesion strength for PI on aluminum, and a lesser decrease for the icephobic coating as well. This decrease is due to the increased occurrence of cohesive failures. Between -10°C and -18°C, there is a transition from adhesive failures to more cohesive failures for aluminum and PI, as shown previously [24]. The same transition can be seen for the icephobic coating (see supplementary materials). At NTNU on the other hand, there was only one occurrence of cohesive failure for aluminum surfaces, which occurred at -10°C when using the VST. These observations indicate that the transition to cohesive failures and the occurrence and impact of non-adhesive failures depends on the ice adhesion test method and ice type. For BWI on the icephobic coating tested at NTNU, there is a slight increase of ice adhesion strength with temperature. The varying effect of temperature on the ice adhesion strength for the different configurations substantiate the difficulty in predicting the dependence of ice adhesion strength on temperature, as reported previously [17].

In general terms, this study shows that there are large differences between different laboratories, and that the differences do not seem to be systematic. It seems that for higher ice adhesion strengths, the difference between different ice adhesion tests and ice types increases. It follows that more tests with a larger range of ice adhesion values are needed to explore this relation more fully.

As two different ice types were tested at AMIL, the similar trend from Rønneberg et al. [40] can be seen in that BWI has a lower ice adhesion strength than PI for aluminum. However, for the icephobic coating, the ice adhesion strength for both ice types is very similar. As a result, it may be that the difference in ice adhesion strength between different types of ice depends on whether the tested surface is defined as a low adhesion surface.

When comparing the results from AMIL and NTNU, some general comments about different ice adhesion measurement set-ups can be made. At low ice adhesion, the two test methods give similar results, while the VST seems to give larger deviations than the CAT methodology. However, the VST is easier to implement, and has a slightly lower standard deviation for low ice adhesion surfaces with BWI. An alternative might be the lap shear test, as studied recently [57], although no comparison can be made between this new test method and the ones presented in this study.

Lastly, some additional sources of error present in the experiments reported here must be mentioned. For the tests performed at NTNU, the ice adhesion tests were performed ex situ and the ice samples and tested surfaces were moved between the freezer to the testing apparatus. Especially the tests performed at -10°C were subject to a long transport between two different laboratories, and to account for this thermal variation, a polystyrene container was used. The effect of this container compared to the shorter transport at room temperature for the tests performed at -18°C cannot be determined exactly. However, despite the transport which was assumed detrimental for ice adhesion, the NTNU VST methods yields higher ice adhesion for both coatings, compared to AMIL results where the experiments were performed in situ. This observation may indicate that the transportation did not significantly affect the ice adhesion.

The ice sample size differed between AMIL and NTNU, with an iced area of about 1100 mm$^2$ at AMIL while only 594 mm$^2$ at NTNU. While at AMIL, the ice sample covers the entire tested surface as seen in Figures 1a and 1b, at NTNU the ice sample is situated at a part of the tested surface only, as seen in Figure 3. The fact that the ice sample at NTNU is smaller compared to the surface structure, especially for aluminum, may be a factor in the much higher standard deviation seen for the aluminum samples from NTNU than the icephobic coating.

# Concluding remarks

In this study, the ice adhesion strength of two different surfaces has been tested at two laboratories with different ice adhesion test methods and two types of accreted ice. Despite the differences between the laboratories, the experiments capture the overall trends in which the ice adhesion strength surprisingly decreases from -10°C to -18°C for aluminum and is almost independent of temperature for a commercial icephobic coating. For BWI, the NTNU VST method systematically yields higher ice adhesion strength than AMIL CAT method. The standard deviations were approximately constant when testing PI at AMIL and seems to scale with the absolute value of ice adhesion at NTNU. The VST has higher deviations than CAT methodology for high ice adhesion values, but are more similar when testing low ice adhesion surfaces.

The experiments in this study were performed with a focus on keeping the conditions similar, both within each lab and between AMIL and NTNU. However, the results still show significant differences and variations. As a result, more data from several more laboratory facilities are needed as well as more tests within each laboratory facility. Furthermore, the study provides further evidence that the ice formation is a key parameter in predicting the ice adhesion on different surfaces, as well as for the investigation of the mechanism of the ice detachment from different surfaces and the occurrence of cohesive failures during ice adhesion testing.

## Acknowledgements

The authors gratefully acknowledge the financial support from the Norwegian Research Council FRINATEK project Towards Design of Super-Low Ice Adhesion Surfaces (SLICE, 250990) and from the PETROMAKS2 project Durable Arctic Icephobic Materials (AIM, 255507).

The authors declare no conflicts of interests.

# Interlaboratory study of ice adhesion using different techniques
# Supplementary materials

Sigrid Rønneberg, Yizhi Zhuo, Caroline Laforte, Jianying He, Zhiliang Zhang

## Contents

### S1. Experimental protocol

*Table S2 Experimental protocol used for the interlaboratory study. See Experimental section for more info on procedures.*

| Facility | Ice type | Surface # | Temperature | Repetitions | Icing time | Waiting time |
|---|---|---|---|---|---|---|
| **AMIL** | Precipitation ice | Aluminum | -10 | 6 | 33min | 1h |
| **AMIL** | Precipitation ice | Coating | -10 | 6 | 33min | 1h |
| **AMIL** | Precipitation ice | Aluminum | -18 | 6 | 33min | 1h |
| **AMIL** | Precipitation ice | Coating | -18 | 6 | 33min | 1h |
| **AMIL** | Bulk water ice | Aluminum | -10 | 6 | 3h | 15min |
| **AMIL** | Bulk water ice | Coating | -10 | 6 | 3h | 15min |
| **AMIL** | Bulk water ice | Aluminum | -18 | 6 | 3h | 15min |
| **AMIL** | Bulk water ice | Coating | -18 | 6 | 3h | 15min |
| **NTNU** | Bulk water ice | Aluminum | -10 | 5 | 3h | 15min |
| **NTNU** | Bulk water ice | Coating | -10 | 5 | 3h | 15min |
| **NTNU** | Bulk water ice | Aluminum | -18 | 5 | 3h | 15min |
| **NTNU** | Bulk water ice | Coating | -18 | 5 | 3h | 15min |

**Notes**

- Temperature relates to both freezing temperature and testing temperature
- Initial temperature of both surfaces and water was room temperature for bulk water ice
- All surfaces were only tested once

## S2. All experimental results

*Table S3 Experimental results from the ice adhesion tests for all 66 samples.*

| Surface | | Aluminum | Aluminum | Coating | Coating |
|---|---|---|---|---|---|
| Temperature | | $T_{air}$ = -10°C | $T_{air}$ = -18°C | $T_{air}$ = -10°C | $T_{air}$ = -18°C |
| **AMIL, precipitation ice** | 1 | 727 | 81 | 265 | 62 |
| | 2 | 741 | 79 | 320 | 81 |
| | 3 | 782 | 84 | 387 | 90 |
| | 4 | 788 | 85 | 346 | 59 |
| | 5 | 774 | 81 | 344 | 83 |
| | 6 | 589 | 86 | 380 | 90 |
| | **Mean** | **734** | **83** | **340** | **78** |
| | **SD** | **75** | **3** | **44** | **14** |
| | | **10 %** | **4 %** | **13 %** | **18 %** |
| **AMIL, bulk water ice** | 1 | 343 | 118 | 269 | 139 |
| | 2 | 346 | 70 | 315 | 119 |
| | 3 | 281 | 39 | 318 | 39 |
| | 4 | 332 | 113 | 193 | 17 |
| | 5 | | 127 | 294 | 121 |
| | 6 | | 106 | 318 | 72 |
| | **Mean** | **326** | **96** | **285** | **85** |
| | **SD** | **30** | **34** | **49** | **49** |
| | | **9 %** | **36 %** | **17 %** | **58 %** |
| **NTNU, bulk water ice** | 1 | 375 | 118 | 338 | 182 |
| | 2 | 543 | 96 | 467 | 158 |
| | 3 | 405 | 134 | 257 | 143 |
| | 4 | 819 | 119 | 569 | 97 |
| | 5 | 402 | 88 | 332 | 96 |
| | **Mean** | **509** | **111** | **393** | **135** |
| | **SD** | **185** | **19** | **124** | **38** |
| | | **36 %** | **17 %** | **32 %** | **28 %** |

## S3. Ice formation

The formation of bulk water ice is illustrated in Figures 1, 2 and 3 for AMIL and NTNU, respectively. For the generation of precipitation ice, we refer to other publications [1, 2].

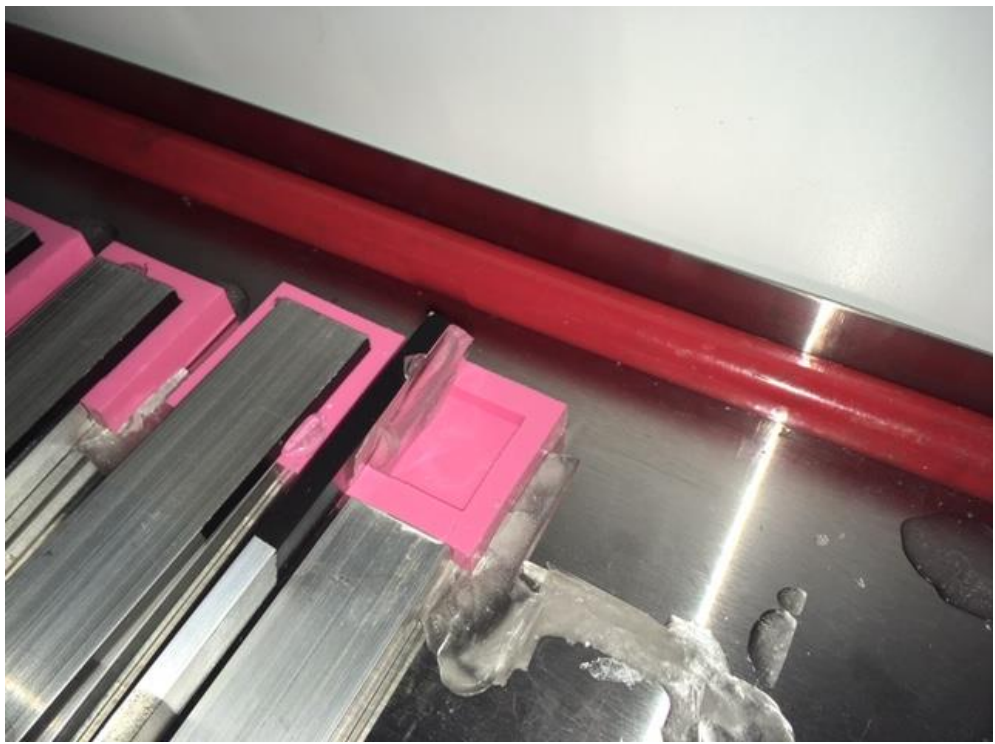

*Figure S1 Formation of bulk water ice at AMIL, same procedure for both temperatures.*

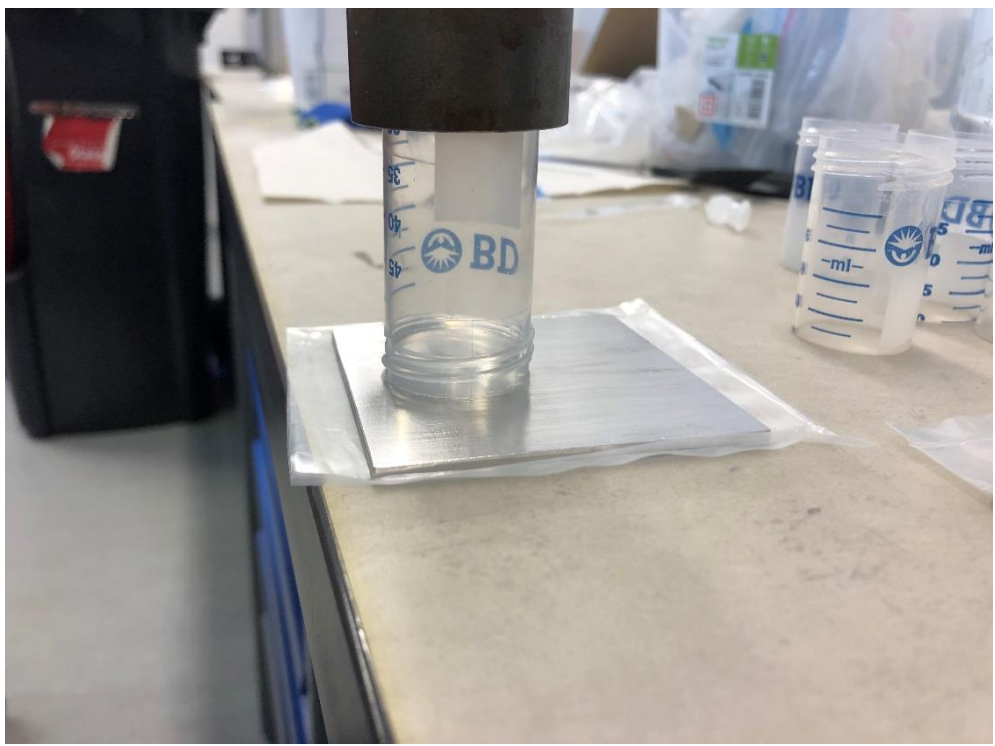

*Figure S2 Formation of bulk water ice on aluminum surface at NTNU. For $T_{air}$ = -18°C, the water was added in room temperature and moved to the freezer. For -10°C, the water insertion was performed in a cold room, otherwise with the same procedure.*

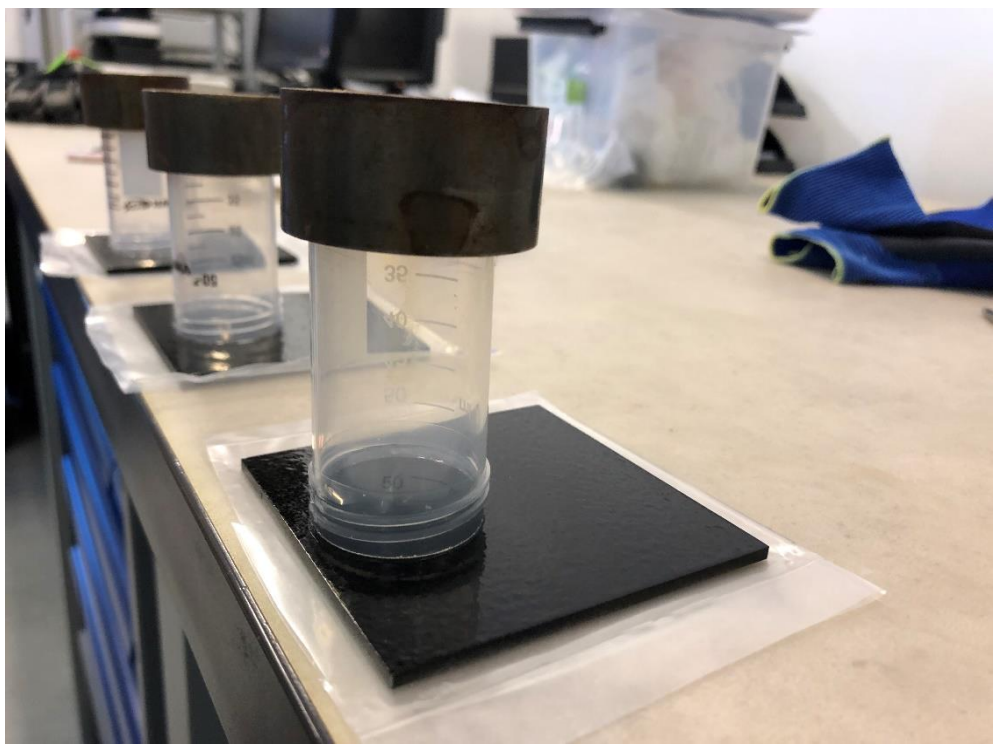

*Figure S3 Formation of bulk water ice on icephobic coating at NTNU, similar to Figure 2.*

## S4. Typical failure modes

Typical failure modes when testing ice adhesion strength can be seen in Figures 4-12. For bulk water ice, the failures were adhesive. For precipitation ice at AMIL, the failures were mostly adhesive at $T_{air}$ = -10°C and cohesive at $T_{air}$ = -18°C.

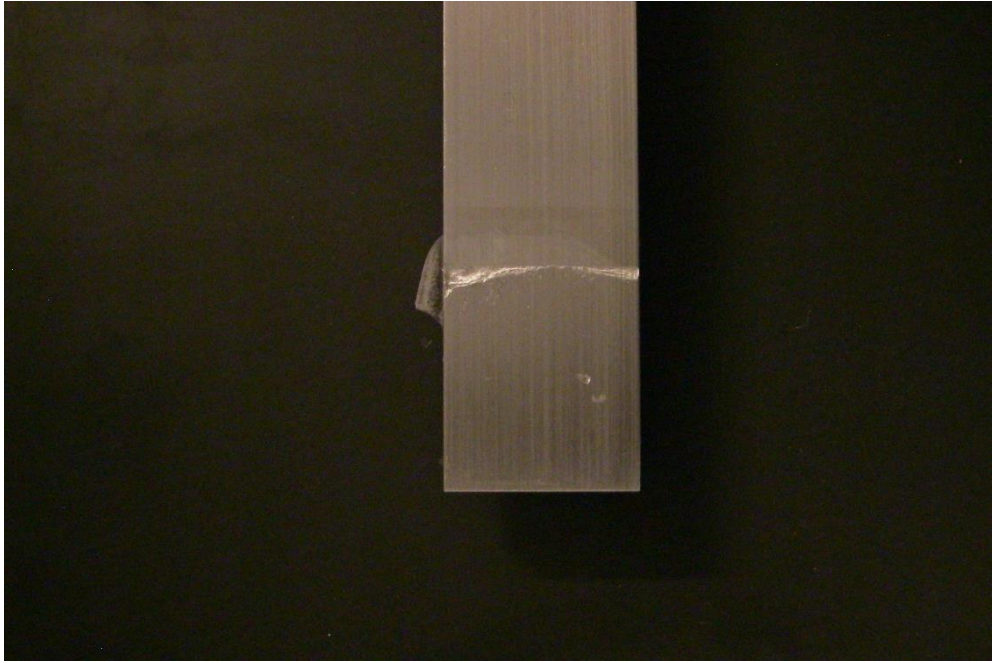

*Figure S4 Typical adhesive failure observed at AMIL for bulk water ice at both temperatures, here for aluminum surface.*

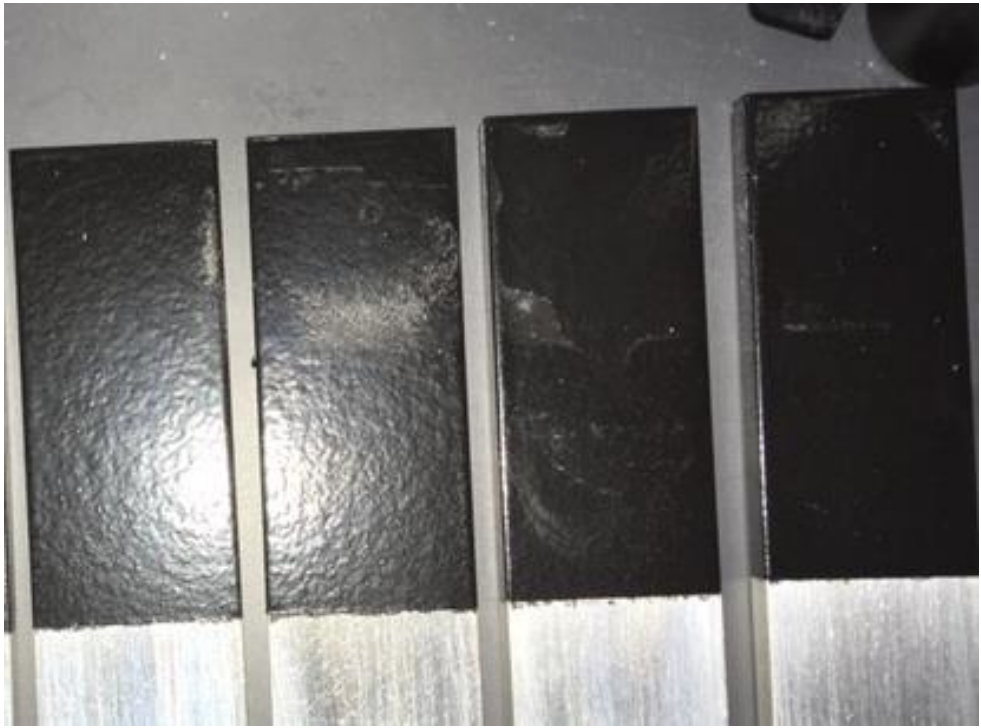

*Figure S5 Typical adhesive failure observed at AMIL for bulk water ice at both temperatures, here for the icephobic coating.*

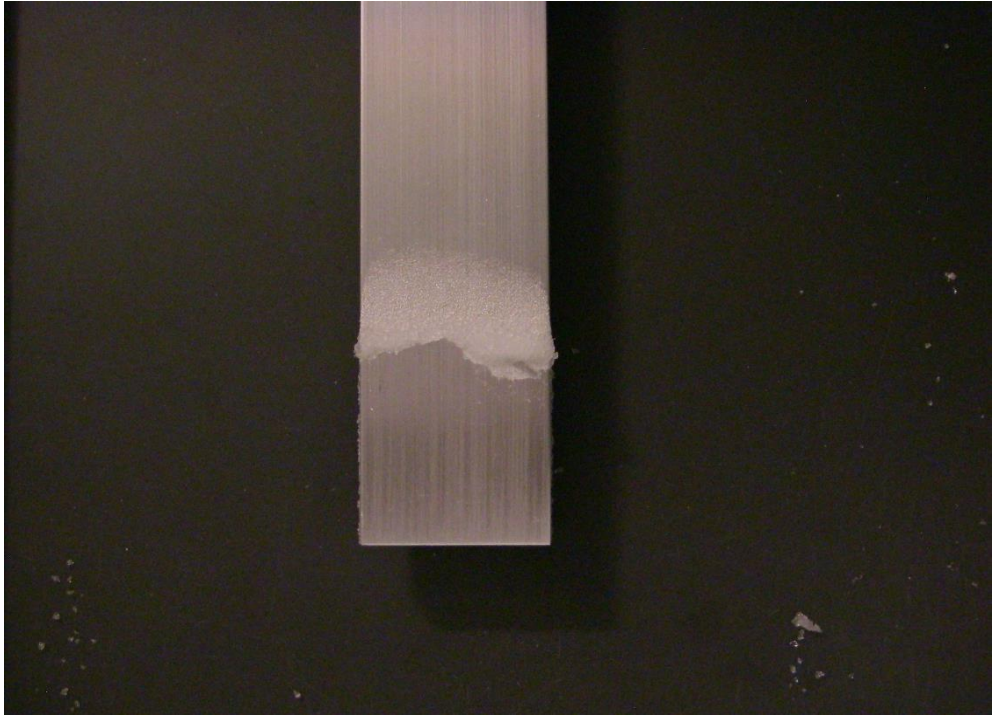

*Figure S6 Adhesive failure observed at AMIL for precipitation ice at $T_{air}$ = -10°C, here for aluminum surface.*

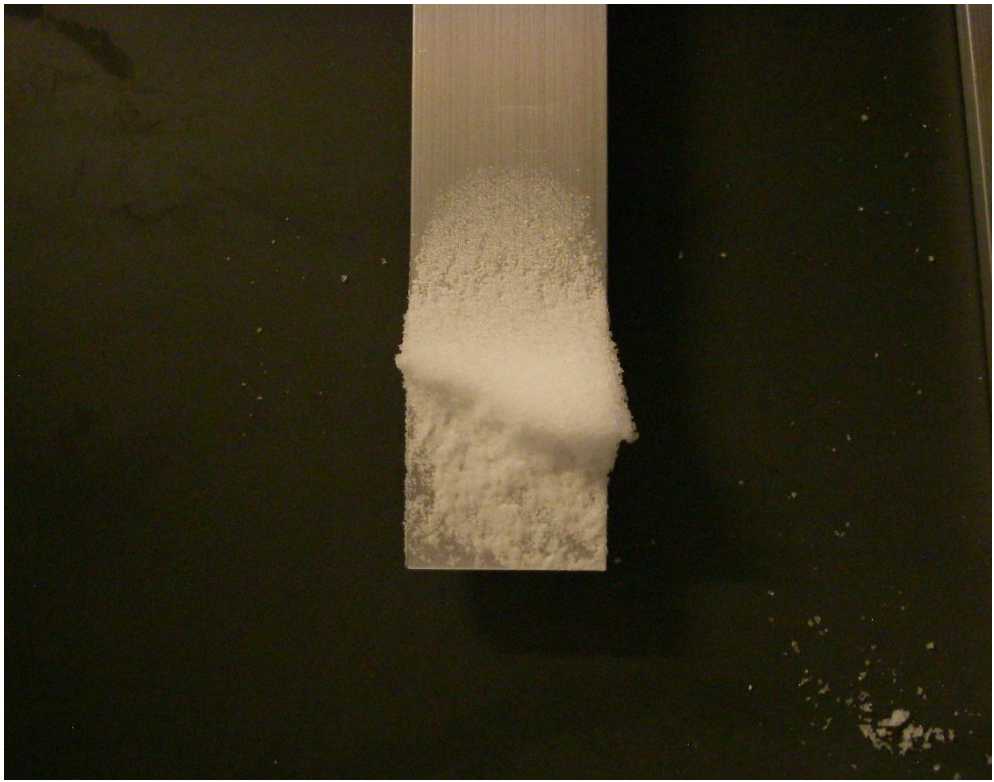

*Figure S7 Cohesive failure observed at AMIL for precipitation ice at $T_{air}$ = -18°C, here for aluminum surface.*

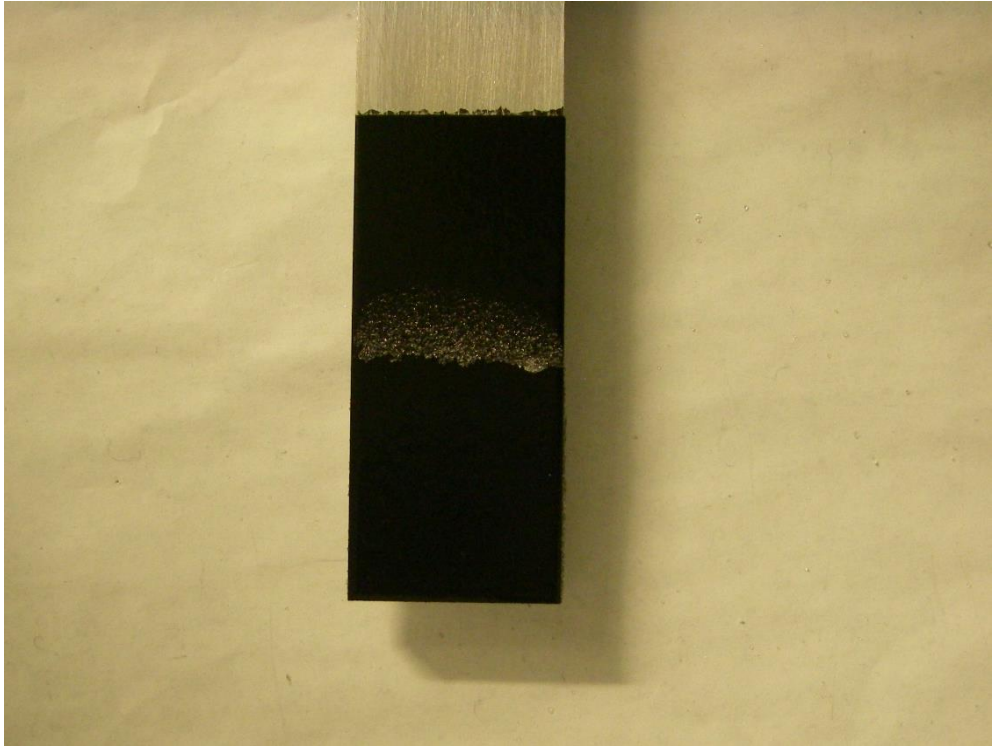

*Figure S8 Adhesive failure observed at AMIL for precipitation ice at $T_{air}$ = -10°C, here for icephobic coating.*

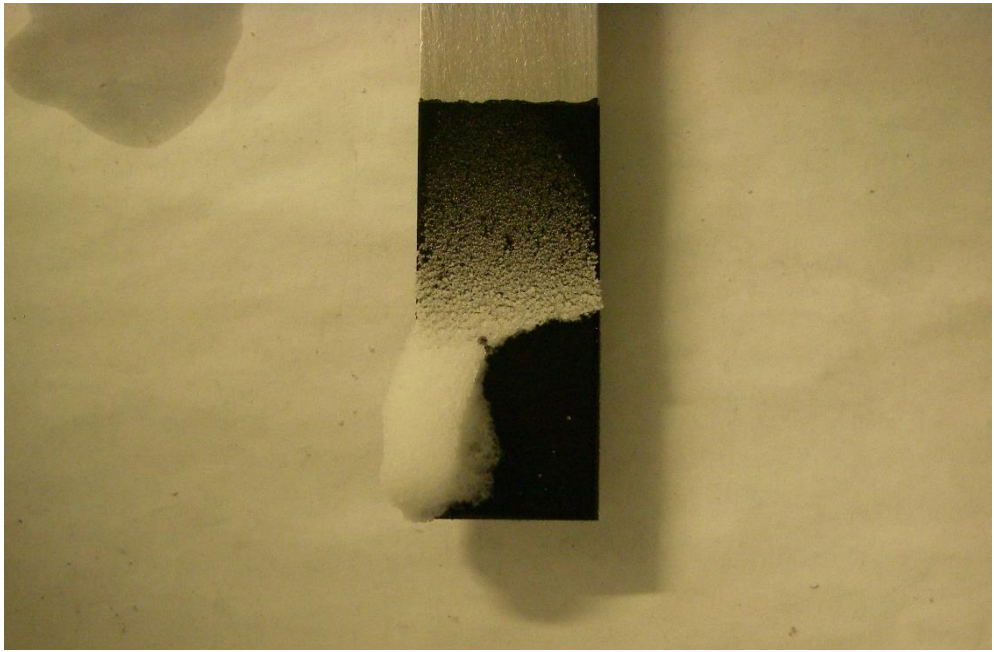

*Figure S9 Cohesive failure observed at AMIL for precipitation ice at $T_{air}$ = -18°C, here for icephobic coating.*

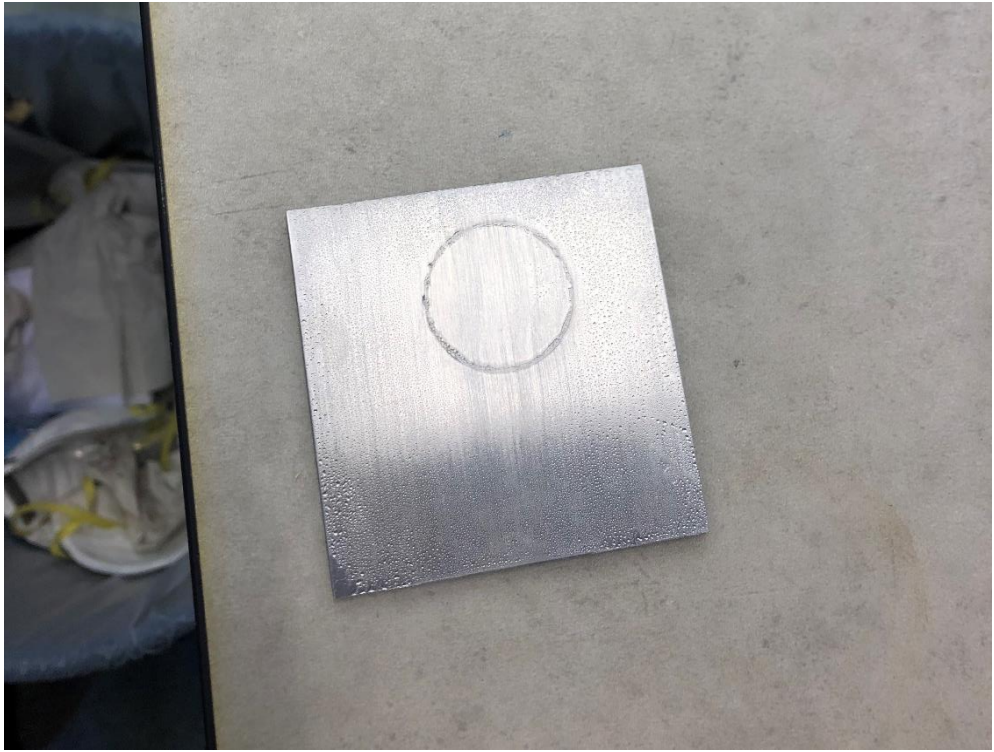

*Figure S10 Typical adhesive failure at ice detachment for tests performed at NTNU. Here for aluminum surface tested at $T_{air}$ = -18°C.*

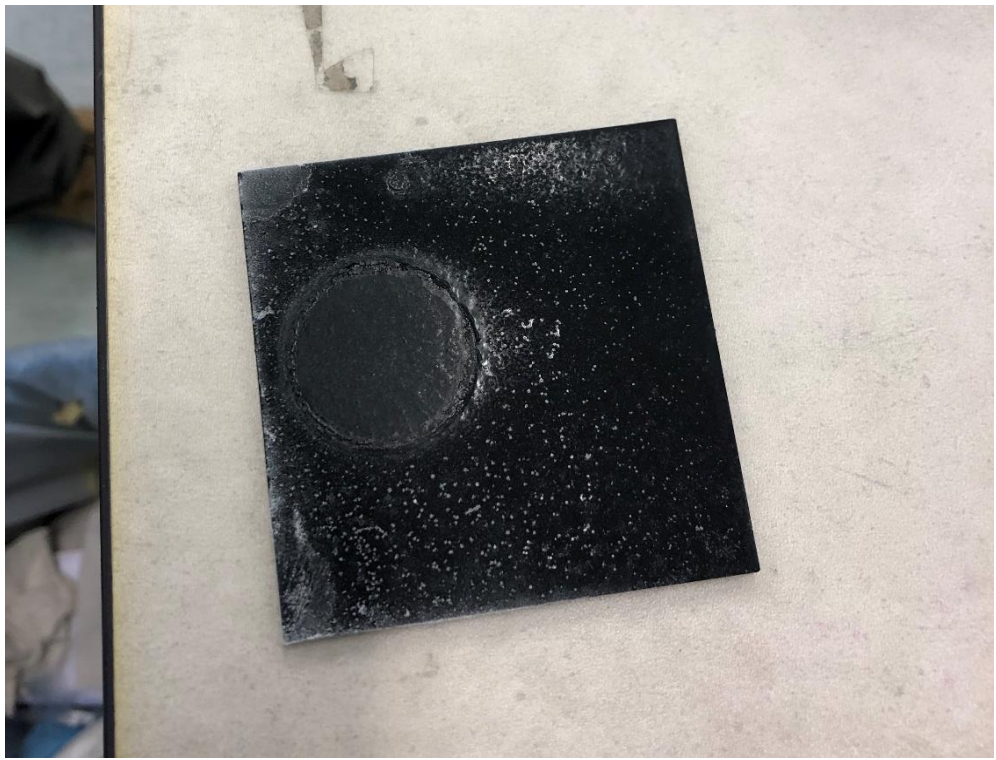

*Figure S11 Typical adhesive failure at ice detachment for tests performed at NTNU. Here for icephobic surface tested at $T_{air}$ = -18°C.*

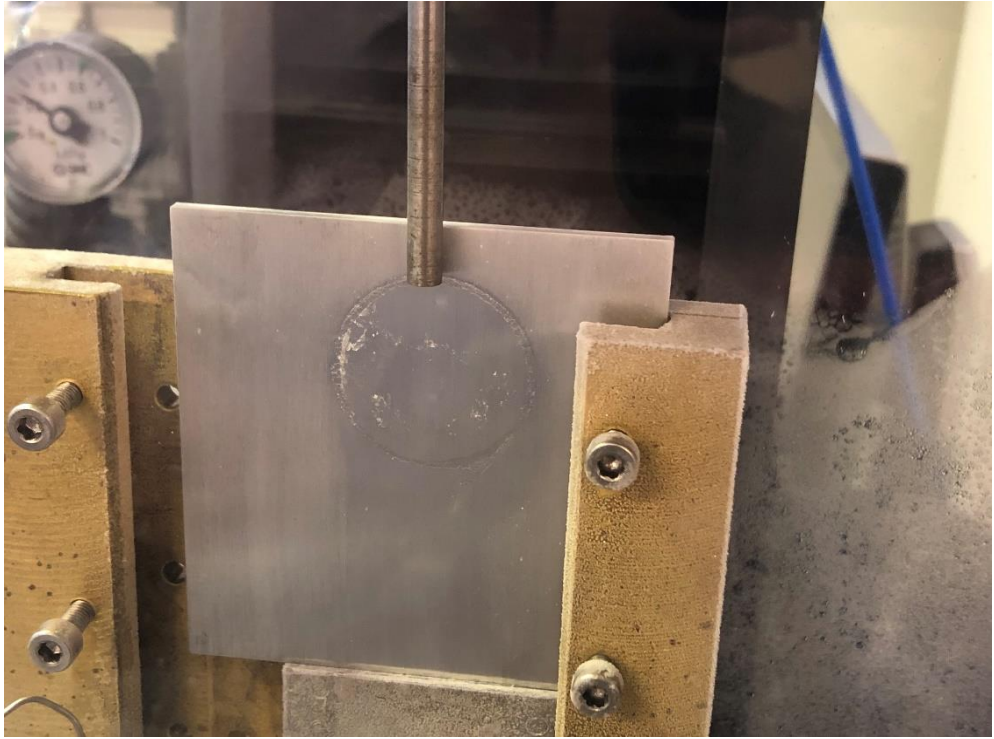

*Figure S12 Picture of the only cohesive failure observed for tests at NTNU. This failure occurred for aluminum surface at $T_{air}$ = -10ºC.*

## S5. Adhesion reduction factor (ARF)

The Adhesion reduction factor (ARF) is defined as the ratio of the ice adhesion strength of a reference material, often aluminum, to the ice adhesion strength of the coating being tested [3]. If the ARF is above 1, the coating has an improved anti-icing behavior. The ARF for the coating tested in this study is shown in Figure 13 for all configurations of ice type and laboratory. The discussion of the ARF is left for a later publication.

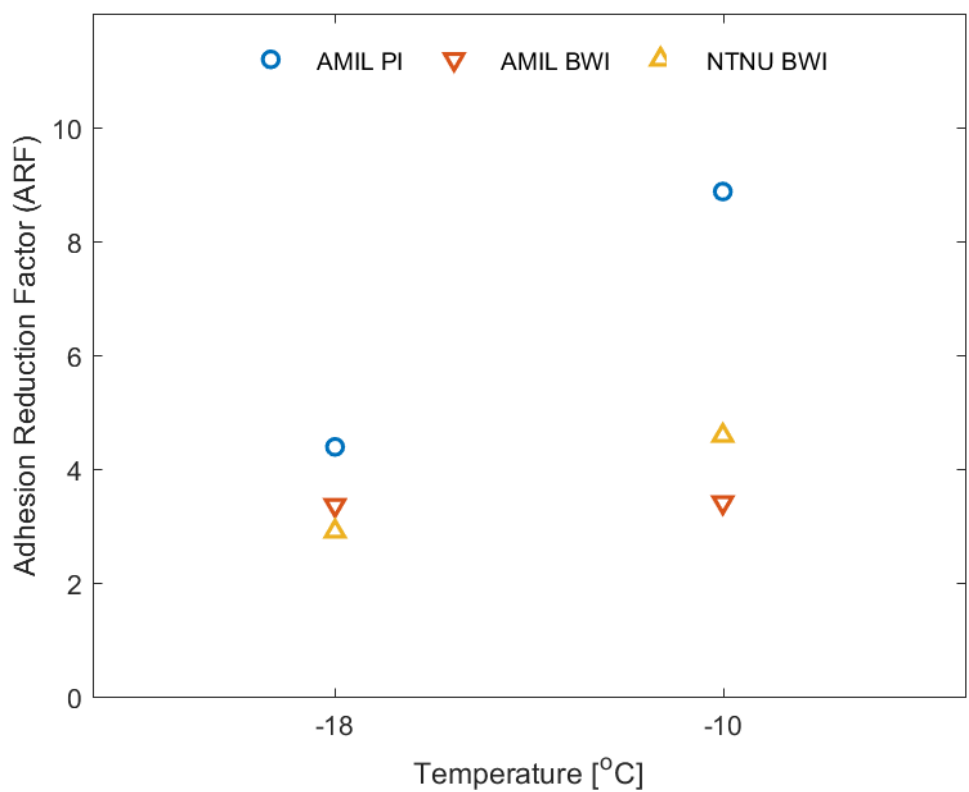

*Figure S13 Overview of ARF for the three ice types for both temperatures.*